\documentclass[conference]{IEEEtran}

\usepackage{amsmath}
\usepackage{graphicx}
\usepackage{makecell}
\usepackage{caption}
\usepackage{subcaption}
\usepackage{multirow}
\usepackage{float}
\usepackage{array}
\usepackage[font=small,labelfont=bf]{caption}
\usepackage{url}

\usepackage{lineno}
\usepackage{colortbl}
\definecolor{Gray}{gray}{0.9}

\usepackage{color,soul}

\graphicspath{{Images/}}

\makeatletter
\def\ps@IEEEtitlepagestyle{%
  \def\@oddfoot{\mycopyrightnotice}%
  \def\@oddhead{\hbox{}\@IEEEheaderstyle\leftmark\hfil\thepage}\relax
  \def\@evenhead{\@IEEEheaderstyle\thepage\hfil\leftmark\hbox{}}\relax
  \def\@evenfoot{}%
}
\def\mycopyrightnotice{%
  \begin{minipage}{\textwidth}
  \centering \scriptsize
  \copyright 2023 IEEE.  Personal use of this material is permitted.  Permission from IEEE must be obtained for all other uses, in any current or future media, including reprinting/republishing this material for advertising or promotional purposes, creating new collective works, for resale or redistribution to servers or lists, or reuse of any copyrighted component of this work in other works. Presented at IEEE BioCAS 2023. Final version available at: https://doi.org/10.1109/BioCAS58349.2023.10388660
  \end{minipage}
}
\makeatother

\begin{document}

\makeatletter
\def\ps@IEEEtitlepagestyle{%
  \def\@oddhead{}%
  \def\@evenhead{}%
  \def\@oddfoot{\mycopyrightnotice\hfill\thepage}%
  \def\@evenfoot{\hfill\thepage}%
}
\makeatother

\makeatletter
\def\ps@headings{%
  \def\@oddhead{}
  \def\@evenhead{}%
  \def\@oddfoot{\hfill\thepage\hfill}
  \def\@evenfoot{\hfill\thepage\hfill}%
}
\makeatother

\pagestyle{headings}

%
\title{FMCW Radar and Vital Activity-Based Subject Detection for Multi-Subject Vital Sign Monitoring}
\title{Design and Measurement of Truly Non-Contact mmWave Radar Based Vital Monitoring System for Multiple patients}
\title{Design and Measurement of Truly Non-Contact mmWave Radar Based Vital Monitoring System for Multiple patients in Mass Monitoring Scenario}
\title{Design and Prototyping of Multi-Patient Truly Non-Contact mmWave Radar Based Heart Rate and Breath Rate Measurement System}
\title{Design and Measurement of Truly Non-Contact mmWave Radar Based Heart Rate and Breath Rate Measurement System for Multiple patients}
\title{Design and Measurements of Multi-Patient Truly Non-Contact mmWave Radar Based Heart Rate and Breath Rate Measurement System}
\title{Design and Measurements of Multi-Patient Truly Non-Contact mmWave Radar Based Heart Rate and Breath Rate Monitoring System}
\title{Design and Experimentation with non-Contact mmWave FMCW Radar for Multi-Patient Vital Sign}
\title{Non-Contact mmWave FMCW Radar for Multi-Patient Vital Sign: Design and Measurements}
\title{Design and Measurements of non-Contact mmWave FMCW Radar for Multi-Patient Vital Sign}
\title{Design and Measurements of mmWave FMCW Radar Based Non-Contact Multi-Patient Heart Rate and Breath Rate Monitoring System}

 \author{Jewel Benny$\textsuperscript{a}$, Pranjal Mahajan$\textsuperscript{a}$, 
 Srayan Sankar Chatterjee$\textsuperscript{a}$,
   Mohd Wajid$\textsuperscript{b}$, Abhishek Srivastava$\textsuperscript{a}$\\
 \textit{$\textsuperscript{a}$Centre for VLSI and Embedded Systems Technology (CVEST), IIIT Hyderabad, India}\\
 \textit{$\textsuperscript{b}$Department of Electronics Engineering, Z.H.C.E.T. Aligarh Muslim University, India}
 }

\maketitle



\begin{abstract}
Recent developments in mmWave radar technologies have enabled the truly 
non-contact heart-rate (HR) and breath-rate (BR) measurement approaches, which provides a great ease in patient monitoring. Additionally, these technologies also provide opportunities to simultaneously detect HR and BR of multiple patients, which has become increasingly important for efficient mass monitoring scenarios. 
In this work, a frequency modulated continuous wave (FMCW) mmWave radar based truly non-contact multiple patient HR and BR monitoring system has been presented. Furthermore, a novel approach is also proposed, which combines multiple processing methods using a least squares solution to improve measurement accuracy, generalization, and handle measurement error.
The proposed system has been developed using Texas Instruments' FMCW radar and experimental results with multiple subjects are also presented, which show  \textgreater 97\% and \textgreater 93\% accuracy in the measured BR and HR values, respectively.
\end{abstract}

\IEEEpeerreviewmaketitle

\section{Introduction}

\begin{figure}
    \centering
    \includegraphics[width=0.85\linewidth]{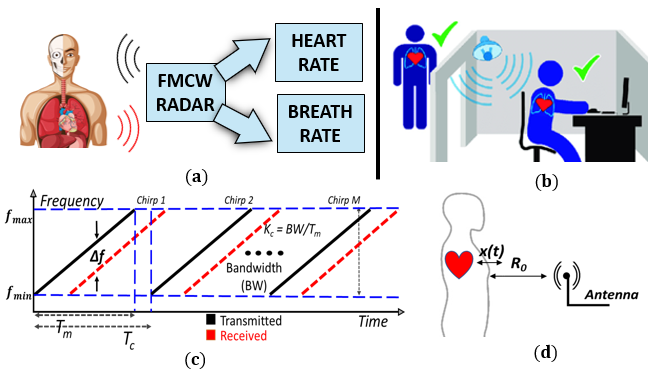}
    \caption{(\textit{a}) Heart rate and breath rate estimation using FMCW radar (\textit{b}) Multi-subject HR \& BR mesurement (\textit{c}) F-T plot of FMCW chirps (\textit{d}) Chest displacement due to breathing and heart beat}
    \label{intro_fig}
     \vspace{-5mm}
\end{figure}

\IEEEPARstart{M}{illimeter} wave (mmWave) technologies such as frequency modulated continuous wave (FMCW) radar near 77 GHz spectrum has gained great attention as a non-contact alternative to wearable devices for mass monitoring of vital signs. As depicted in Fig. \ref{intro_fig}(\textit{a}) and Fig. \ref{intro_fig}(\textit{b}), these radars can be used for mass monitoring vital signs such as breath-rate (BR) and heart-rate (HR). 
FMCW radars are emerging as superior alternatives due to their ability to measure micro displacements associated with respiration ($\sim$1-12 mm) and heartbeat ($\sim$0.01-0.5 mm) \cite{info1,cw_to_fmcw}. 
While previous studies \cite{cw_to_fmcw}-\cite{comp3} have demonstrated HR/BR monitoring using FMCW radars, they primarily focused on single-subject scenarios. Multi-subject measurement was shown in \cite{comp4}, however, it lacks simultaneous measurement of multiple subjects and require prior knowledge of subject azimuth to optimize signal-to-noise ratio (SNR).

In this work, we present 
\textit{(i)} design of a mmWave FMCW radar-based system with high accuracy to simultaneously measure BR/HR of multiple patients, 
\textit{(ii)} a combination of multiple estimation methods using a least squares solution \cite{least_squares} for improved measurement accuracy, and
\textit{(iii)} measurement results of the proposed system prototype to validate the multi-patient monitoring scenario. 
The paper is structured as follows: Section \ref{sec2} provides a brief theoretical background on FMCW radars. Section \ref{sec3} explains the proposed method for estimating vital signs of multiple patients using FMCW radar in detail. Section \ref{sec4} discusses the experimental setup and results, and finally, Section \ref{sec5} concludes the paper.


\vspace{-2mm}
\section{Background of FMCW Radars}
\label{sec2}
\vspace{-1mm}
As shown in Fig. \ref{intro_fig} (\textit{c}), a chirp is a frequency modulated signal, where its instantaneous frequency changes continuously. 
Expression of a real-valued linear chirp ($y_{TX}(t)$) transmitted by FMCW radar is given by,
\begin{equation}
\label{eq_chirp}
y_{TX}(t) = A_{TX}\cos(2\pi f_{min}t + K_c\pi t^2) 
\end{equation}
where $A_{TX}$ is the signal amplitude, $f_{min}$ is the initial frequency of the chirp at $t=0$ and $K_c$ is the chirp rate \cite{info2,eq_cit1,eq_cit2}. 
Electromagnetic waves travel at the speed of light, $c$, therefore, distance, $R$, traveled by the wave in a time interval $t_d/2$ is given by $R = \frac{ct_d}{2}$, where $t_d$ is the round trip delay. 
Expression for the chirp received by the radar ($y_{RX}(t)$) is given by \cite{eq_cit1,eq_cit2},
\begin{equation}
\label{eq_chirp_rx}
y_{RX}(t) = A_{RX}\cos\big(2\pi f_{min}(t-t_d) + K_c\pi (t-t_d)^2\big).
\end{equation}
Mixing of transmitted and received chirps followed by low pass filtering gives intermediate frequency (IF) signal ($y_{MX}(t)$) as given by,
\begin{equation}
\label{eq_chirp_if}
    y_{MX}(t) = A\cos(2\pi f_{min}t_d+2\pi K_ctt_d-\pi K_ct_d^2).
\end{equation}
Since, $R$ is few meters, $t_d$ is very small, hence $\pi K_c t_d^2 \approx 0$ \cite{info2}. The IF beat signal ($y_{IF,real}(t)$) can be given by, 
\begin{equation}
\label{eq_if_beat}
    y_{IF,real}(t) \approx A\cos\big(2\pi f_bt+\frac{4\pi R}{\lambda_{max}}\big)
\end{equation}
where $f_b = K_ct_d$ and $\lambda_{max} = c/f_{min}$. A quadrature phase shifted version ($ y_{IF,imag}(t)$) of Eq. (\ref{eq_if_beat}) is also an associated signal, which can be given by, 
\begin{equation}
\label{eq_if_imag}
    y_{IF,imag}(t) \approx   A\sin\big(2\pi f_bt+\frac{4\pi R}{\lambda_{max}}\big). 
\end{equation}
The combined mixed IF signal ($y_{IF,real}(t) + j.y_{IF,imag}(t)$) is represented as complex exponential as given by,

\vspace{-4mm}
\begin{equation}
    y_{IF}(t)\approx  A\exp(j(2\pi f_bt + \phi_R)),
    \label{beat_signal}
    \vspace{-1mm}
\end{equation}
where $\phi_R = \frac{4\pi R}{\lambda_{max}}$ is the phase shift in the mixer output due to an object present at a distance $R$ from the radar.

\section{Proposed Method for Breath-rate and Heart-rate Estimation of Multiple Patients}
\label{sec3}

\begin{figure}
    \centering
    \includegraphics[width=0.85\linewidth]{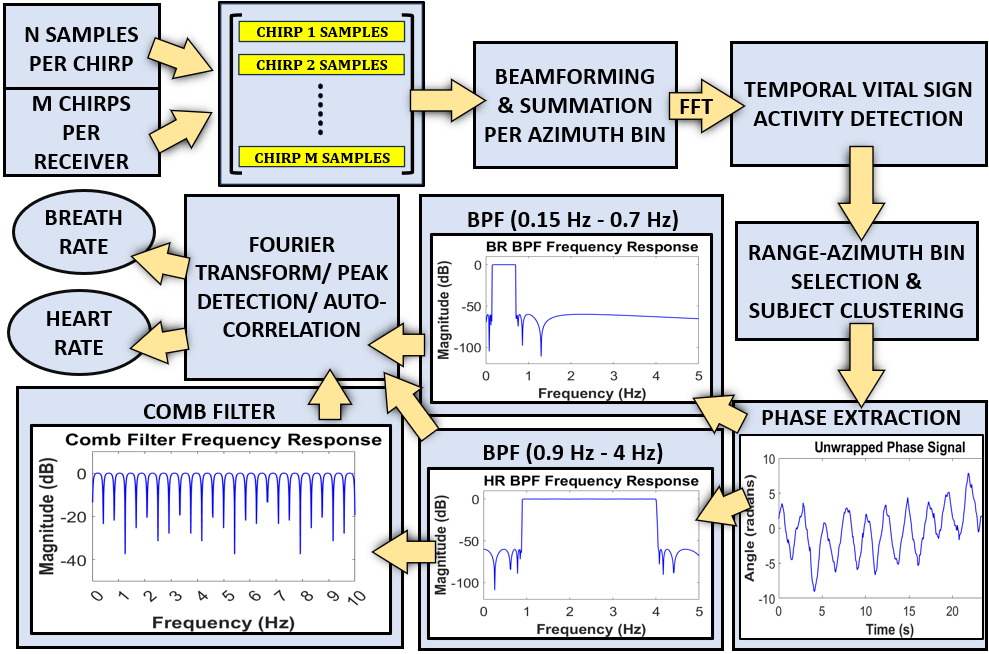}
    \caption{Signal processing flow to measure HR \& BR}
    \label{signal_processing}
    \vspace*{-1.3\baselineskip}
\end{figure}

\begin{figure}
    \centering    \includegraphics[width=8.4cm,height=5.2cm]{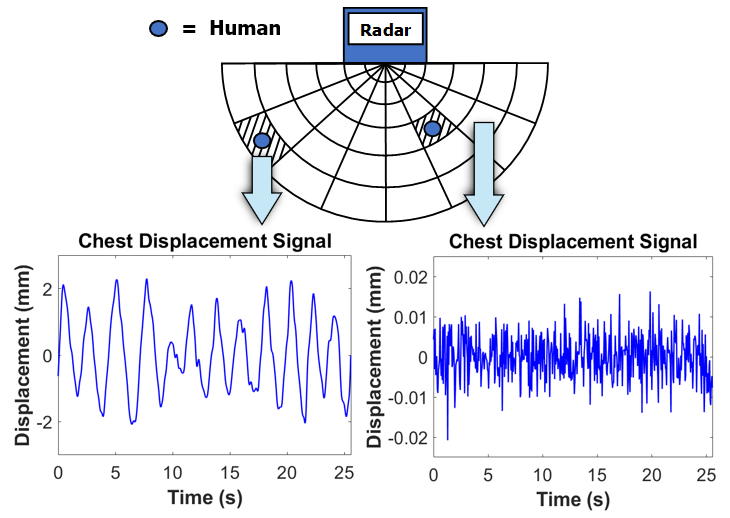}
    \caption{Chest displacement signals extracted from the phase signal of a range-azimuth bin with a human subject (left) and without a human subject (right). }
    \label{range_azimuth_bins}
    \vspace*{-1.3\baselineskip}
\end{figure}

The proposed method employs a multi-receiver radar, beamforming weights, and advanced signal processing techniques to achieve accurate localization and estimation of BR/HR for multiple patients within a room. Fig. \ref{signal_processing} illustrates the proposed signal processing flow for estimating BR/HR.
FMCW radars are better suited for BR/HR estimation since the chest displacements are in the millimeter order; this results in a significant change in the beat signal's phase depending on the chest wall's displacement \cite{info2}. As depicted in Fig. \ref{intro_fig}(\textit{d}), the phase signal ($R(t)$) can be captured as a function of the chest displacement ($x(t)$) with respect to time, $R(t) = R_0 + x(t)$
\cite{info2}.
The phase shift ($\phi_R(t)$) of the mixed signal is a function of time and given by,
\begin{equation}
\label{eq_phir}
\phi_R(t) = \frac{4\pi R(t)}{\lambda_{max}}.
\end{equation}
Simultaneous range and azimuth resolution of the multiple subjects can be done using a multi-receiver system and beamforming weights \cite{beamform, comp5}. Since the beat signals in Eq. (\ref{beat_signal}) will contain frequencies and phase from all possible range-azimuth bins (RABs), therefore it is critical to extract the phase from the specific RAB at which the subjects are located. For a specific azimuth direction $\gamma$, the beat signal ($y_{IF,\gamma}(t)$) can be computed by applying beamforming as given by, 
\begin{equation}
\label{eq_beam_IF}
y_{IF,\gamma}(t) = \sum\limits_{i=1}^{no. of RX} y_{IF}(t)_{i}w_k^i,
\end{equation}
where $ y_{IF}(t)_{i}$ is the beat signal corresponding to the $i^\text{th}$ receiver and $w_k^i$ are beamforming weights chosen such that the interference from other azimuth directions is minimised.  For the specific azimuth, we can estimate the range by taking the Fast Fourier Transform (FFT) of $y_{IF,\gamma}(t)$ with frequency axis scaled to the range axis, called the \textit{Range FFT} \cite{info2}. The magnitude spectrum of the \textit{Range FFT} for a particular azimuth direction, $\gamma$, gives a peak at $R_0^\gamma = \frac{cf_b}{2K_c}$; and phase at $R_0^\gamma$ can also be extracted from this \textit{Range FFT}. When evaluated at $R_0^\gamma+x^\gamma(t)$, the \textit{Range FFT} of $y_{IF,\gamma}(t)$ is given by,
\vspace{-0.1mm}
$$
\small
Y_{\gamma}\big(R_0^\gamma+x^\gamma(t)\big) = \mathcal{F}\Big(y_{IF,\gamma}(t)\Big)\big|_{R=R_0^\gamma+x^\gamma(t) \qquad \qquad\qquad\qquad\qquad \qquad} 
\normalsize
$$
\begin{equation}
\small
       \quad \approx \mathcal{F}\Big(A\exp(j(2\pi ft))\Big)\big|_{R=R_0^\gamma}.\exp\Big(\frac{4\pi [R_0^\gamma+x^\gamma(t)]}{-j\lambda_{max}}\Big).
       \label{equation9}
\normalsize
\end{equation}
This means that the phase of $Y_{\gamma}$ changes with the subject range. If a single chirp is transmitted, the range estimated using magnitude \textit{Range FFT} is almost constant for a given subject. Consider when transmitting the sequence of M number of chirp pulses with chirp duration $T_m$ and inter-chirp interval $T_c$, as shown in Fig. \ref{intro_fig} (\textit{c});
 and assume estimated range from magnitude of \textit{Range FFT} is $R_{0,m}^\gamma$ for the $m^\text{th}$ chirp  \& azimuth $\gamma$ , the Eq. (\ref{equation9}) is re-written as, 
 $$ 
 \small
Y_{\gamma}\big(R_{0,m}^\gamma+x_m^\gamma(t)\big) \qquad \qquad \qquad \qquad \qquad  \qquad  \qquad  \qquad  \qquad 
 \normalsize
 $$
\begin{equation}
\label{eq_range_fft_if}
\small
    \approx \mathcal{F}\Big(A\exp(j2\pi ft)\Big)\Big|_{R=R_{0,m}^\gamma}.\exp\Big(\frac{4\pi [R_{0,m}^\gamma+x_m^\gamma(t)]}{-j\lambda_{max}}\Big).    
\normalsize
\end{equation}
After computing \textit{Range FFT} for all possible RABs, the proposed vital sign activity detection and estimation algorithm is applied. Fig. \ref{range_azimuth_bins} shows the extracted $R(t)$ signals from RABs with and without human following the proposed flow of Fig. \ref{signal_processing}. Details of the proposed procedure is presented in the following subsections.

 \begin{figure*}
    \centering \includegraphics[width=0.9\linewidth]{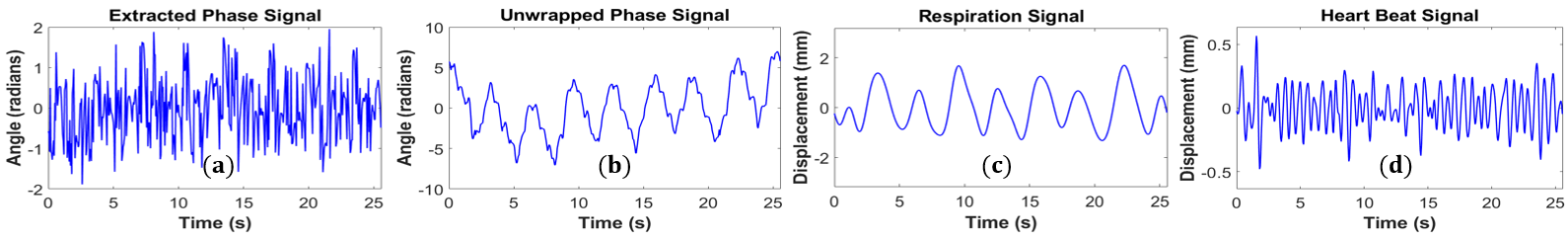}
    \caption{(\textit{a}) Extracted phase signal (\textit{b}) Unwrapped phase signal (\textit{c}) Extracted breathing signal (\textit{d}) Extracted heartbeat signal}
    \label{phase_sigs}
    \vspace*{-1.3\baselineskip}
\end{figure*}

\subsection{{Detecting Temporal Activity}}
A sequence of M chirps is transmitted; each chirp produces a unique \textit{Range FFT} for a RAB. The chest wall's millimeter order displacements cause changes in the beat signal's phase, resulting in variations in the \textit{Range FFT}s across chirps in the RABs. The stationary clutters are easily eliminated since the \textit{Range FFT}s across the chirps are not changing for the clutter bin.
Also, for each azimuth direction, the average of \textit{Range FFT} magnitude is computed across M chirps to create a combined \textit{Range FFT} magnitude, thereby, estimating the more accurate range of the patient. 
After checking the presence of temporal activity in all bins, we combine them into a matrix called the \textit{Vital Activity Map}, where each row corresponds to a range bin and each column corresponds to an azimuth bin. Then, a threshold-based elimination process removes RABs with minimal or no vital sign activity. This process creates a sparse matrix that roughly represents the RABs corresponding to the location of the patients. 
\subsection{{Phase Extraction for Detected Patients}}
Taking $Y_{\gamma}\big(R_{0,m}^\gamma+x_m^\gamma(t)\big)$ across M chirps over the selected RABs, then calculating its phase for every chirp gives the required phase signal, whose sampling interval is $T_c$. 
This phase signal is further processed by subtracting with its mean followed by phase unwrapping \cite{phase_unwrap}. The phase signal is unwrapped beyond $[-\pi, \pi]$ to ensure that the difference between two consecutive phase samples is less than $\pi$. 

The frequency band of human BR and HR signals are distinct and typically in the range of 3-36 per minute and 48-120 per minute, respectively \cite{info2}. So, the unwrapped phase signal is passed through specific bandpass filters to obtain the clean breathing signal $\phi_{br}(m)$ and heartbeat signal $\phi_{hr}(m)$. Fig. \ref{phase_sigs} shows, (\textit{a}) the extracted phase signal, (\textit{b}) the unwrapped phase signal, and (\textit{c}) the extracted breathing signal and (\textit{d}) the extracted heartbeat signal. Signals $\phi_{br}(m)$ and $\phi_{hr}(m)$ are processed further to obtain BR and HR.
\begin{figure*}
    \centering
    \includegraphics[width=0.93\textwidth,height=7.5cm]{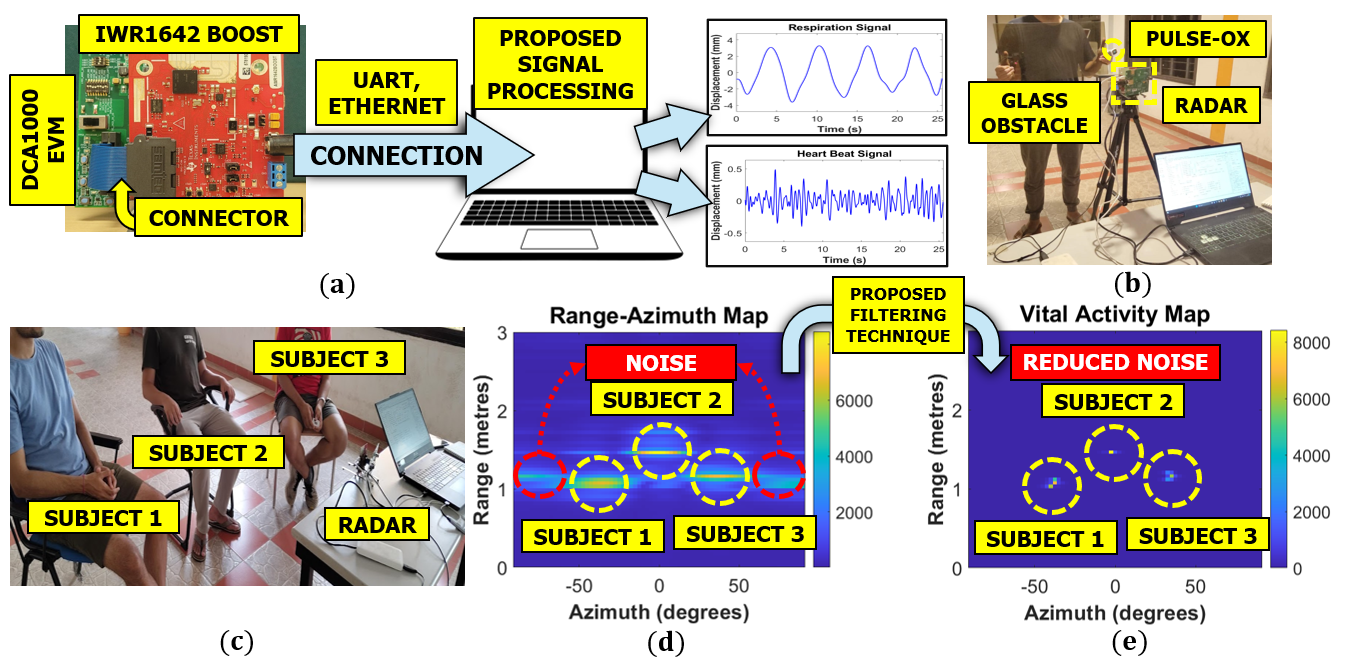}
    \caption{(\textit{a}) The overall experiment setup (\textit{b}) Single-subject HR \& BR measurement (\textit{c}) Multi-subject HR \& BR measurement (\textit{d}) Range-azimuth map (\textit{e}) Vital activity map after clutter removal}
    \label{fig_4}
    \vspace*{-0.8\baselineskip}
\end{figure*}
\begin{table*}[h!]
    \centering
    \fontsize{6.51pt}{6pt}
    \caption{Measurement Results}
    \vspace{-1.3mm}
    \setlength{\tabcolsep}{2.5pt}
    \resizebox{0.9\linewidth}{!}{
    \begin{tabular}{| c | c | c | c c c c c c | c c c c c c c c c |}
        \hline 
            \multirow{2}{*}{\thead{Exp. \\ No.}} & \multirow{2}{*}{\thead{Patient \\ No.}} & \multirow{2}{*}{\thead{Range (m) \\ and  \\ Azimuth (\textdegree)}} & 
            \multicolumn{6}{c|}{\thead{Breath Rate (per min)}} & 
            \multicolumn{9}{c|}{\thead{Heart Rate (per min)}} \\
            \cline{4-9}\cline{10-18}
            &  & & \thead{$br_f$} & \thead{$br_a$} & \thead{$br_p$} & \thead{$br$} & \thead{Manual \\ Counting} & \thead{Error (\%)} & \thead{$hr_f$} & \thead{$hr_a$} & \thead{$hr_p$} & \thead{$hr_{fc}$} & \thead{$hr_{ac}$} & \thead{$hr_{pc}$} & \thead{$hr$} & \thead{Reading From \\ Beuerer PO 30} & \thead{Error\\(\%)} \\
        \hline
        \rowcolor{Gray}
        1 & $1 $ & $5, 0$\textdegree & $ 24.44 $ & $ 26.09 $ & $ 25.78  $ & $ 25.40 $ & $ 25.78 $ & $ 1.47 $ & $ 117.00 $ & $ 103.48 $ & $ 106.77 $ & $ 125.98 $ & $ 101.56 $ & $ 120.39 $ & $ 124.24 $ & $ 110 $ & $ 12.95 $\\ 
        
        2 & $1 $ & $0.5, -15$\textdegree & $ 10.01 $ & $ 9.06 $ & $ 9.38 $ & $ 9.52 $ & $ 9.37 $ & $ 1.60 $ & $ 69.41 $ & $ 78.93 $ & $ 78.15 $ & $ 76.68 $ & $ 67.30 $ & $ 76.22 $ & $ 69.79 $ & $ 70 $ & $ 0.30 $\\ 
        \rowcolor{Gray}
         & $1 $ & $1, -30$\textdegree & $ 10.54 $ & $ 11.41 $ & $ 11.36 $ & $ 11.10 $ & $ 10.55 $ & $ 5.21 $ & $ 72.55 $ & $ 66.05 $ & $ 63.09 $ & $ 71.48 $ & $ 75.91 $ & $ 62.80 $ & $ 67.09 $ & $ 66 $ & $ 1.65 $\\
        \rowcolor{Gray}
         \multirow{-2}{*}{3}& $2 $ & $1, 30$\textdegree & $ 18.75 $ & $ 17.87 $ & $ 18.75 $ & $ 18.59 $ & $ 18.75 $ & $ 0.85 $ & $ 82.68 $ & $ 81.33 $ & $ 78.21 $ & $ 76.78 $ & $ 80.12 $ & $ 86.45 $ & $ 79.35 $ & $ 87 $ & $ 8.79 $\\
        
        \multirow{2}{*}{4} & $1 $ & $2, -45$\textdegree & $ 8.92 $ & $ 8.95 $ & $ 9.26 $ & $ 9.09 $ & $ 9.37 $ & $ 2.99 $ & $ 84.72 $ & $ 75.86 $ & $ 80.12 $ & $ 80.33 $ & $ 84.67 $ & $ 76.95 $ & $ 78.83 $ & $ 71 $ & $ 11.03 $\\ 
         & $2 $ & $1.5, 60$\textdegree & $ 13.76 $ & $ 14.01 $ & $ 12.42 $ & $ 13.09 $ & $ 12.89 $ & $ 1.55 $ & $ 73.04 $ & $ 86.23 $ & $ 87.41 $ & $ 83.77 $ & $ 79.14 $ & $ 83.20 $ & $ 74.30 $ & $ 77 $ & $ 3.51 $\\
        \rowcolor{Gray}
         & $1 $ & $1, -45$\textdegree & $ 11.95 $ & $ 12.47 $ & $ 12.21 $ & $ 12.17 $ & $ 12.30 $ & $ 1.06 $ & $ 75.36 $ & $ 77.33 $ & $ 71.43 $ & $ 77.33 $ & $ 80.96 $ & $ 70.54 $ & $ 72.45 $ & $ 83 $ & $ 12.71 $\\ 
        \rowcolor{Gray}
        & $2 $ & $1.5, 0$\textdegree & $ 23.06 $ & $ 25.36 $ & $  22.68 $ & $ 23.23 $ & $ 23.44 $ & $ 0.90 $ & $ 145.67 $ & $ 135.20 $ & $ 131.92 $ & $ 144.44 $ & $ 147.05 $ & $ 142.89 $ & $ 137.61 $ & $ 130 $ & $ 5.85 $\\
        \rowcolor{Gray}
         \multirow{-3}{*}{5} & $3 $ & $1.2, 45$\textdegree & $ 22.01 $ & $ 22.08 $ & $ 21.16 $ & $ 21.56 $ & $ 21.68 $ & $ 0.55 $ & $ 103.71 $ & $ 106.37 $ & $ 110.74 $ & $ 109.65 $ & $ 106.03 $ & $ 105.35 $ & $ 102.16 $ & $ 102 $ & $ 0.16 $\\
        
        \multirow{3}{*}{6} & $1 $ & $2, -60$\textdegree & $ 16.31 $ & $ 17.06 $ & $ 16.83 $ & $ 16.70 $ & $ 16.40 $ & $ 1.83 $ & $ 81.35 $ & $ 85.08 $ & $ 76.11 $ & $ 74.09 $ & $ 80.67 $ & $ 85.91 $ & $ 68.04 $ & $ 76 $ & $ 10.47 $\\ 
         & $2 $ & $2, 15$\textdegree & $ 11.36 $ & $ 10.23 $ & $ 11.33 $ & $ 11.15 $ & $ 10.55 $ & $ 5.69 $ & $ 84.36 $ & $ 84.08 $ & $ 80.65 $ & $ 82.34 $ & $ 82.36 $ & $ 79.18 $ & $ 77.11 $ & $ 72 $ & $ 7.1 $\\ 
         & $3 $ & $1.5, 60$\textdegree & $ 18.30 $ & $ 17.66 $ & $ 17.48 $ & $ 17.75 $ & $ 17.58 $ & $ 0.97 $ & $ 99.03 $ & $ 95.15 $ & $ 92.66 $ & $ 95.40 $ & $ 90.73 $ & $ 92.55 $ & $ 91.80 $ & $ 88 $ & $ 4.32 $\\
        \hline
        \multicolumn{3}{|c|}{} & & & & \multicolumn{2}{c}{\thead{Average BR Error (\%)}} & \thead{2.06} & & & & & & & \multicolumn{2}{c}{\thead{Average HR Error (\%)}} & \thead{6.57}\\
        \hline
    \end{tabular}}
    
    \label{measurement_results}
    \vspace*{-1.2\baselineskip}
\end{table*}


\begin{table}
    \fontsize{7pt}{6pt}
    \caption{Comparison Table}
    \setlength{\tabcolsep}{3.4pt}
    \centering
    \resizebox{0.9\linewidth}{!}{
    \begin{tabular}{| c | c | c | c | c | c |}
    \hline
         &  \thead{Type \\ of \\ Radar} & \thead{BR \\ Accuracy \\ (\%)} & \thead{HR \\ Accuracy \\ (\%)} & \thead{Maximum \\ Range \\ Reported (m)} & \thead{Number of \\ simultaneous  \\ patients reported} \\
    \hline
    \rowcolor{Gray}
    \textbf{Our work} & FMCW & 97.94 &  93.43 & 5 & 3 \\
    \cite{cw_to_fmcw} & CW & 88.42 & 87.22 & 1.5 & 1 \\
    \rowcolor{Gray}
    \cite{info2} & FMCW & 94 & 80 & 1.7 & 1 \\
    \cite{comp2} & FMCW & 95.15 & 82.45 &  2 & 1 \\
    \rowcolor{Gray}
    \cite{comp3} & FMCW & $>$ 93 & $>$ 93 & 0.7 & 1 \\
    \cite{comp4} & FMCW & 93 & 95 & 0.5 & 2 \\
    \rowcolor{Gray}
    \cite {comp5} & FMCW & - &  - & 1  & 2 \\
    \cite{comp6} & FMCW & - &  - & 1 & 2 \\
    \hline
    \end{tabular}}

    \label{comparison_table}
    \vspace*{-1.8\baselineskip}
\end{table}
\vspace{-3mm}
\subsection{{BR and HR Estimation}} BR and HR are estimated using 
Fourier transform, auto-correlation, and peak detection, followed by an optimal linear combination of the estimates. 
 \subsubsection{Fourier Transform} The spectrum of $\phi_{br}(m)$ is analyzed to determine the breath rate $br_f$ by identifying the frequency with the highest peak. Similarly, the spectrum of $\phi_{hr}(m)$  is analyzed to determine the heart rate $hr_f$  by averaging the frequencies of the $l$ highest peaks (in this case $l=6$). 
\subsubsection{Auto-Correlation}
The auto-correlation of M length sequence $\phi(m)$ is given by \cite{auto_corr},
\begin{equation}
    R_{\phi\phi}[n] = \frac{1}{M}\sum\limits_{m=-\infty}^{\infty}\phi[m]\phi[n+m].
\end{equation}
The auto-correlation function of a periodic signal has the property that it retains the period of the original signal, with a peak at $n=0$. 
The auto-correlation functions of $\phi_{br}(m)$ and $\phi_{hr}(m)$ are calculated and the time inverse of the second peak is taken as $br_a$ and $hr_a$ respectively. 
\subsubsection{Peak Detection}
The number of peaks in signals $\phi_{br}(m)$ and $\phi_{hr}(m)$ above a particular threshold are calculated and divided by the time duration of the signals. The result is taken as $br_p$ and $hr_p$, respectively. 

It is critical to suppress the respiration harmonics in the signal $\phi_{hr}(m)$ as it can overshadow the actual heartbeat frequencies. Therefore, $\phi_{hr}(m)$ is also passed through a comb filter \cite{respiration_harmonics}.
After passing $\phi_{hr}(m)$ through the comb filter and again applying the three signal processing techniques processing methods, as discussed above, we obtain three more measurements, namely, $hr_{fc}$, $hr_{ac}$, $hr_{pc}$. 

\subsubsection{Optimal Linear Combination}
 Linear combinations of the estimates will further improve the accuracy. Hence, a linear combination of $br_f$, $br_a$ and $br_p$ for BR, and $hr_f$, $hr_a$, $hr_p$, $hr_{fc}$, $hr_{ac}$ and $hr_{pc}$ for HR are taken as the final vital rates,
\vspace{-0.2mm}
\begin{equation}
    br = c_fbr_f+c_abr_a+c_pbr_p
\end{equation}
and
\begin{equation}
    hr = \Big(\begin{gathered}d_fhr_f+d_ahr_a+d_phr_p+\\ \,\,d_{fc}hr_{fc}+d_{ac}hr_{ac}+d_{pc}hr_{pc}\end{gathered}\Big),
\end{equation}
where $c_f, c_a, c_p, d_f, d_a$, $d_p$, $d_{fc}$, $d_{ac}$ and $d_{pc}$ are scalars chosen to minimise the least square error between estimated values and ground truth values. Consider that $P$ number of experiments have been performed with ground truth values of HR and BR are in the column vector $\text{\textbf{b}}$ and $\text{\textbf{h}}$ (each of order $P\times1$), respectively. If $\text{\textbf{c}} = [c_f \,\,\, c_a \,\,\, c_p]^T$ and $\text{\textbf{d}} = [d_f \,\,\, d_a \,\,\, d_p \,\,\, d_{fc} \,\,\, d_{ac} \,\,\, d_{pc}]^T$, the ground truth values for BR and HR are given by the column vectors $\text{\textbf{b}}$ and $\text{\textbf{h}}$, the measurements corresponding to  $br_f$, $br_a$ and $br_p$ is given by matrix $\text{\textbf{B}}$ of order $P\times3$, and the measurements corresponding to $hr_f$, $hr_a$, $hr_p$, $hr_{fc}$, $hr_{ac}$ and $hr_{pc}$ is given by matrix $\text{\textbf{H}}$ of order $P\times6$. The least squares solution is given by,
\begin{equation}
    \text{\textbf{c}} = (\text{\textbf{B}}^T\text{\textbf{B}})^{-1}\text{\textbf{B}}^T\text{\textbf{b}}
\end{equation}
\begin{equation}
        \text{\textbf{d}} = (\text{\textbf{H}}^T\text{\textbf{H}})^{-1}\text{\textbf{H}}^T\text{\textbf{h}.}
\end{equation}
In our study, we consider P=50 across different experiments.
\vspace{-2mm}
\section{Experimental Setup and Results}
\label{sec4}


Fig. \ref{fig_4}(\textit{a}) depicts the experimental setup used for measuring BR/HR using TI-IWR1642BOOST radar module (77-81 GHz) \cite{1642} along with TI DCA1000 EVM \cite{DCA} for acquiring the in-phase and quadrature-phase data corresponding to the IF signal. mmWave Studio \cite{mmwave_studio} by TI is used to control the radar module and acquire the signals. The acquired IF signals are  transferred to a computer for further processing in MATLAB and BR/HR extraction. 
The radar uses 2 TX and 4 RX antennas for azimuth resolution via beam-forming. By sequentially transmitting chirps, the TX antennas achieve time division multiplexing, creating a virtual RX array of 8 antennas, which provides an angle resolution of 15 degrees \cite{comp5}.
Each TX-RX pair of the proposed system in Fig. \ref{fig_4}(\textit{a}) is configured to record 512 chirps with $T_c = 50$  ms, resulting in an observation time of 25.6 seconds. This window length was chosen to balance accuracy in generating a \textit{Vital Activity Map} and processing latency. 


Fig. \ref{fig_4}(\textit{b}) shows a single subject standing near the radar with a glass obstacle for BR and HR detection wearing a pulse oximeter \cite{ref_sensor} for validating the measured results. Fig. \ref{fig_4}(\textit{c}) shows three subjects near the radar who were detected in the range-azimuth map shown in Fig. \ref{fig_4}(\textit{d}). As shown in Fig. \ref{fig_4}(\textit{d}), the surrounding objects are also captured in the range-azimuth map which can be considered as noise. This noise is filtered by the proposed temporal vital activity detection, which generates the \textit{Vital Activity Map} shown in Fig. \ref{fig_4}(\textit{e}) demonstrating the area with maximum signs of activity. Further processing was done to obtain their vital signs reported in Table \ref{measurement_results}.
As seen in Table \ref{measurement_results}, an accuracy of 97.94\% for BR and 93.43\% for HR was achieved, demonstrating the effectiveness of the system in accurately measuring BR and HR. The accuracy decreased as the range of the subject increased due to the decrease in SNR received by the radar as range increases \cite{snr_eq}. 
The system accurately measures HR and BR up to a distance of 5 meters from the radar. 
Table \ref{comparison_table} compares this work with similar studies, showing higher accuracy for BR and comparable accuracy for HR measurement. 




\section{Conclusion}
\label{sec5}


In this work, a 77-81 GHz FMCW radar based truly non-contact measurement system for simultaneous detection of HR and BR of multiple subjects has been demonstrated. 
A novel approach is also proposed to improve BR/HR measurement accuracy by combining multiple signal processing methods. The proposed system prototype was developed using TI's FMCW radar and experimental results with multiple subjects are also presented, which show $>$97\% and $>$93\% accuracy in the measured BR and HR values, respectively.

\section*{Acknowledgment}
The authors would like to acknowledge Chips to Startup (C2S) program, Ministry of Electronics and Information Technology (MeitY), Govt. of India, Kohli Center on Intelligent Systems (KCIS) and IHub Mobility IIIT Hyderabad, India for supporting this research.

\ifCLASSOPTIONcaptionsoff
  \newpage
\fi



\end{document}